\documentclass{iaus}
\usepackage{graphicx}

\title[Collision probabilities] 
{Collision probabilities of migrating small bodies and dust particles with planets
}

\author[Sergei I. Ipatov]   
{Sergei I. Ipatov$^{1,2}$}

\affiliation{$^1$Catholic University of America \\ Washington DC, USA \\ email: {\tt siipatov@hotmail.com} \\[\affilskip]
$^2$Space Research Institute, Moscow, Russia}

\pubyear{2009}
\volume{263}  
\pagerange{1--4}
\jname{Icy Bodies of the Solar System}
\editors{D. Lazzaro, D. Prialnik, R. Schulz \& J.A. Fernandez, eds.}
\begin{document}

\maketitle

\begin{abstract}
Probabilities of collisions of migrating small bodies and dust particles
produced by these bodies with planets were studied.
Various Jupiter-family comets,
Halley-type comets, long-period 
comets, trans-Neptunian objects, and asteroids were considered.
The total probability of collisions of any considered body or particle 
with all planets did not exceed 0.2.
The amount of water delivered from outside of Jupiter's orbit to the Earth 
during the formation of the giant planets could exceed the amount
of water in Earth's oceans.
The ratio of the mass of water delivered to a planet by Jupiter-family 
comets or Halley-type comets to the mass of the planet can be greater 
for Mars, Venus, and Mercury, than that for Earth. 

\end{abstract}

\firstsection 
\section{Model of migration of small bodies and dust particles}
In the present paper we discuss the probabilities of collisions of migrating small bodies
and dust particles produced by these bodies with planets.
Ipatov \& Mather (2003, 2004a-b, 2006, 2007), Ipatov et al. (2004),  and several other authors cited in the above papers 
studied the probabilities with Earth, Venus, and Mars.
In our above papers (which can be found on astro-ph 
and http://faculty.cua.edu/ipatov/list-publications.htm), 
the probabilities of the collisions 
were calculated based on the time variations of orbits of bodies and particles
during their dynamical lifetimes (until all bodies or particles reached 
2000 AU from the Sun or collided with the Sun). 
Using  the same time variations,
below we consider also the probabilities of collisions with the giant planets and Mercury.

The orbital evolution of $>$30,000 
bodies with initial orbits close to those of Jupiter-family comets (JFCs), Halley-type 
comets, long-period comets, trans-Neptunian objects, and asteroids in the resonances 3/1 and 5/2 with Jupiter, 
and also of $>$20,000 dust particles produced by these small bodies was integrated  during their dynamical lifetimes. 
 We considered  
the gravitational influence of planets, but omitted the influence 
of Mercury (exclusive for Comet 2P/Encke). 
In about a half of 
calculations of migration of bodies, we used the method by Bulirsh-Stoer (1966) (BULSTO code), 
and in other runs we used a symplectic method (RMVS3 code). The 
integration package of Levison \& Duncan (1994) was used.  
For dust particles, only the BULSTO code was used, and 
the gravitational influence of all planets, the Poynting-Robertson drag, 
radiation pressure, and solar wind drag were taken into account.
The ratio $\beta$ between the radiation pressure force and the gravitational 
force varied from $\le$0.0004 to 0.4. For silicates, such values of $\beta$ 
correspond to particle diameters $d$ between $\ge$1000 and 1 microns; 
$d$ is proportional to 1/$\beta$.

In our calculations, planets were considered as material points, so literal 
collisions did not occur. However, using the algorithm suggested by 
Ipatov (1988) with the correction that takes into account a different 
velocity at different parts of the orbit (Ipatov \& Mather 2003), 
and based on all orbital elements sampled with a 10-500 yr step, we 
calculated the mean probability $P$ of collisions of
migrating objects with a planet. 
The step could be different for different typical dynamical lifetimes of
 particles. We define $P$ 
as $P_{\Sigma}/N$, where $P_{\Sigma}$ is the probability of a collision of 
all $N$ objects  with a planet. 
The probabilities of collisions of bodies and particles
at different $\beta$ 
 with planets 
are presented in Fig. 1.
These probabilities
do not take into account the destruction
of particles in collisions and sublimation,
which can be more important for small particles.
Our runs were made mainly for direct modelling of collisions with the Sun,
but Ipatov \& Mather (2007) 
obtained that the mean probabilities of 
collisions of considered bodies with planets, lifetimes of the bodies that spent millions 
of years in Earth-crossing orbits, and other obtained results 
were practically the same if we consider that bodies disappear 
when  perihelion distance becomes less than the radius of the Sun 
or even several such radii. 


\section{Probabilities of collisions of migrating small bodies 
and dust particles with planets and the Sun}
 
The probability $P_E$ of a collision of a JFC with the Earth exceeded $4\cdot10^{-6}$ 
if initial orbits of bodies were close to those of
several tens of JFCs,
even excluding a few bodies for which the probability 
of a collision of one body with the Earth could be greater than the sum of probabilities for 
thousands of other bodies. The Bulirsh-Stoer method of integration and the symplectic method 
gave similar results. The ratios of probabilities of collisions of JFCs with Venus, Mars, and 
Mercury to the mass of a planet usually were not smaller than those for Earth. 
For most considered bodies, 
the probabilities $P_{Me}$ of collisions of most bodies with Mercury 
(exclusive for Comet 2P/Encke, for which $P_{Me}$$\sim$$P_E$)
were smaller by an order of magnitude than those with Earth or Venus. 

For 
dust particles produced by comets and asteroids, $P_E$  was found to have a maximum ($\sim$0.001-0.02) 
at $0.002\le \beta \le0.01$, i.e., at $d$$\sim$100 $\mu$m (this value of $d$ is in accordance with observational data). 
These maximum values of $P_E$ were usually (exclusive for Comet 2P/Encke) greater at least by an order 
of magnitude than the values for parent comets. Probabilities of collisions of most considered 
particles with Venus did not differ much from those with Earth, 
and those with Mars were 
about an order of magnitude smaller. 
For particles produced by Halley-type comets, 
$P$ was greater for Mercury than for Mars.

Using $P_E=4\cdot10^{-6}$ and assuming that the 
total mass of planetesimals that ever crossed Jupiter's orbit was about 100$m_E$ (Ipatov 
1987, 1993), 
where $m_E$ is the mass of the Earth,  Ipatov \& Mather (2003, 2004a-b, 2007)  concluded
that the total mass of water 
delivered from the feeding zone of the giant planets to the Earth could be about the 
total mass of water in Earth's oceans. 
(Similar conclusion was made by Ipatov (2001) based on other calculations.) 
We supposed that the fraction of water in planetesimals equaled 0.5. 
The ratio of the mass of water delivered to a planet by Jupiter-family comets and 
Halley-type comets to the mass of the planet can be greater for Mars, Venus, and 
Mercury, than that for Earth. This larger mass fraction would result in relatively 
large ancient oceans on Mars and Venus.
The larger value of $P$ for Earth we have calculated compared to those argued by 
Morbidelli et al. (2000) ($P_E\sim(1-3)\cdot10^{-6}$) and Levison et al. (2001) 
($P_E\sim4\cdot10^{-7}$)
 is caused by the fact that 
Levison et al. (2001) did not take into account the gravitational 
influence of the terrestrial planets, and Morbidelli et al. (2000) 
considered low-eccentric initial orbits beyond Jupiter's orbit.
Besides, we considered a larger number of bodies.
The detailed discussion on delivery of water 
and the comparison of our results with the results by other authors were made by
Ipatov \& Mather (2007).
Marov \& Ipatov (2005) discussed the delivery of volatiles to the terrestrial planets.

At the present time, most  authors (e.g., Lunine et al. 2003, Morbidelli et al. 2000, and Petit et al. 2001) consider that 
the outer asteroid belt was the main source of the delivery of water
to the terrestrial planets.
Drake \& Campins (2006) noted that the key argument against an asteroidal source of Earth's water 
is that the O's isotopic composition of Earth's primitive upper mantle matches that of anhydrous 
ordinary chondrites, not hydrous carbonaceous chondrites.
To the discussion of the deuterium/hydrogen paradox of the Earth's oceans
presented by Ipatov \& Mather (2007), we can add that
Genda \& Ikoma (2008) 
showed that D/H in the Earth's ocean increased by a factor of 2-9.



Probabilities of collisions of JFCs
with Saturn typically were smaller by an order of magnitude than those with Jupiter, and
collision probabilities with Uranus and Neptune typically were smaller by 
three orders of magnitude than those with Jupiter.
As only a small fraction of comets collided with all planets during dynamical lifetimes
of comets, the orbital evolution of comets for the considered model 
of material points was close to that for the model when comets collided with a planet 
are removed from integrations. 

 Probabilities of collisions of considered particles and bodies with Jupiter during 
their dynamical lifetimes are smaller than 0.1. They can reach 0.01-0.1 for bodies and 
particles initially moved beyond Jupiter's orbit or in Encke-type orbits. For bodies and 
particles initially moved inside Jupiter's orbit, the probabilities are usually smaller 
than the above range and can be zero. Probabilities of collisions of migrating particles 
(exclusive for trans-Neptunian particles) with other giant planets were usually smaller 
than those with Jupiter. The total probability of collisions of any considered body or 
particle with all planets did not exceed 0.2.

Collisions of planetesimals with a star can cause variations in
observed brightness and spectrum of the star.
In our calculations, the fraction $P_{Sun}$ 
of comets collided with the Sun
during their dynamical lifetimes
was about a few percent. For most JFCs, dynamical lifetimes
are less than 10 Myr, and on average $P_{Sun}$$\sim$0.02.
For dust particles, $P_{Sun}$ depends on $\beta$
and can be considerably greater than for their parent bodies.
For example, for Comet 10P/Tempel 2, $P_{Sun}$$\approx$0.01, and
almost all particles produced by this comet
collide with the Sun (Ipatov \& Mather 2006).




\centerline{}

{\bf Caption to the Figure:}

The probability 
of collisions of dust particles and bodies (during their 
dynamical lifetimes) with  Mercury (subfigure Me),
 Venus (V), Earth (E), Mars 
(Ma), Jupiter (J), Saturn (S), Uranus (U), and Neptune (N) versus $\beta$ (the ratio of the 
radiation to gravitational forces) for particles launched from asteroids 
(ast), trans-Neptunian objects (tno), Comet 2P/Encke at perihelion (2P per), 
Comet 2P/Encke at aphelion (2P aph), Comet 2P/Encke in the middle between 
perihelion and aphelion (2P m), Comet 10P/Tempel 2 (10P), Comet 39P/Oterma (39P), 
long-period comets ($lp$) at eccentricity $e$=0.995 and
perihelion distance  $q$=0.9 AU, and Halley-type comets ($ht$) at 
$e$=0.975 and $q$=0.5 AU (for $lp$ and $ht$ runs, initial inclinations were from 0 to 180$^\circ$, 
and particles were launched near perihelia). If there are two points for the same $\beta$, 
then a plot is drawn via their mean value. Probabilities presented at $\beta$$\sim$10$^{-5}$ are for 
small bodies ($\beta$=0). Probabilities presented only for bodies were calculated for 
initial orbits close to orbits of Comets 9P/Tempel 1 (9P), 22P/Kopff (22P), 
28P/Neujmin (28P), 44P/Reinmuth 2 (44P), and test asteroids from resonances 
3:1 and 5:2 with Jupiter at $e$=0.15 and $i$=10$^\circ$ 
(`ast 3:1' and `ast 5:2'). For series 
$n1$ and $n2$, initial orbits of bodies were close to 10-20 different 
Jupiter-family comets (Ipatov \& Mather 2004b).

\begin{figure}[]
\includegraphics[width=5in]{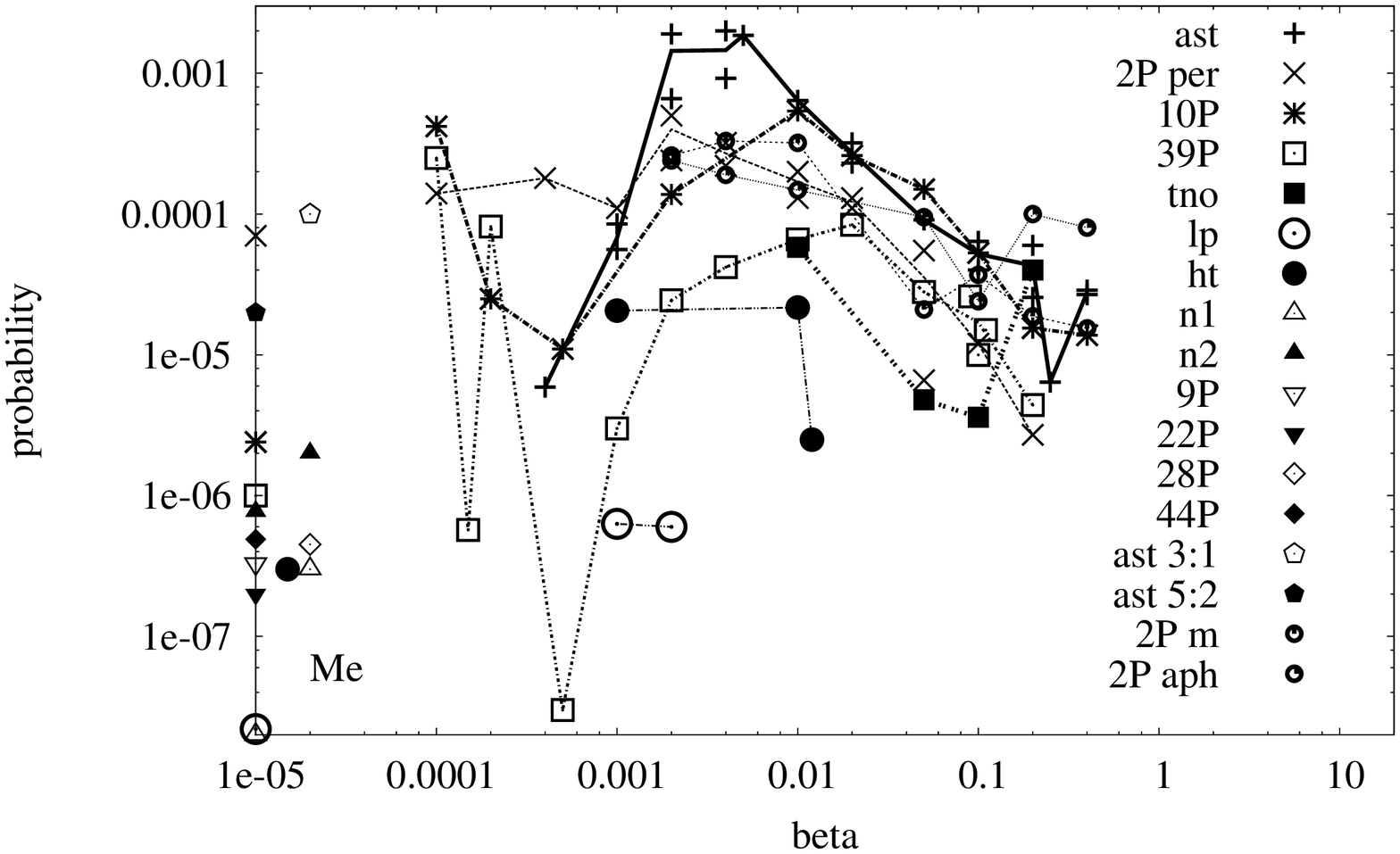}
\includegraphics[width=5in]{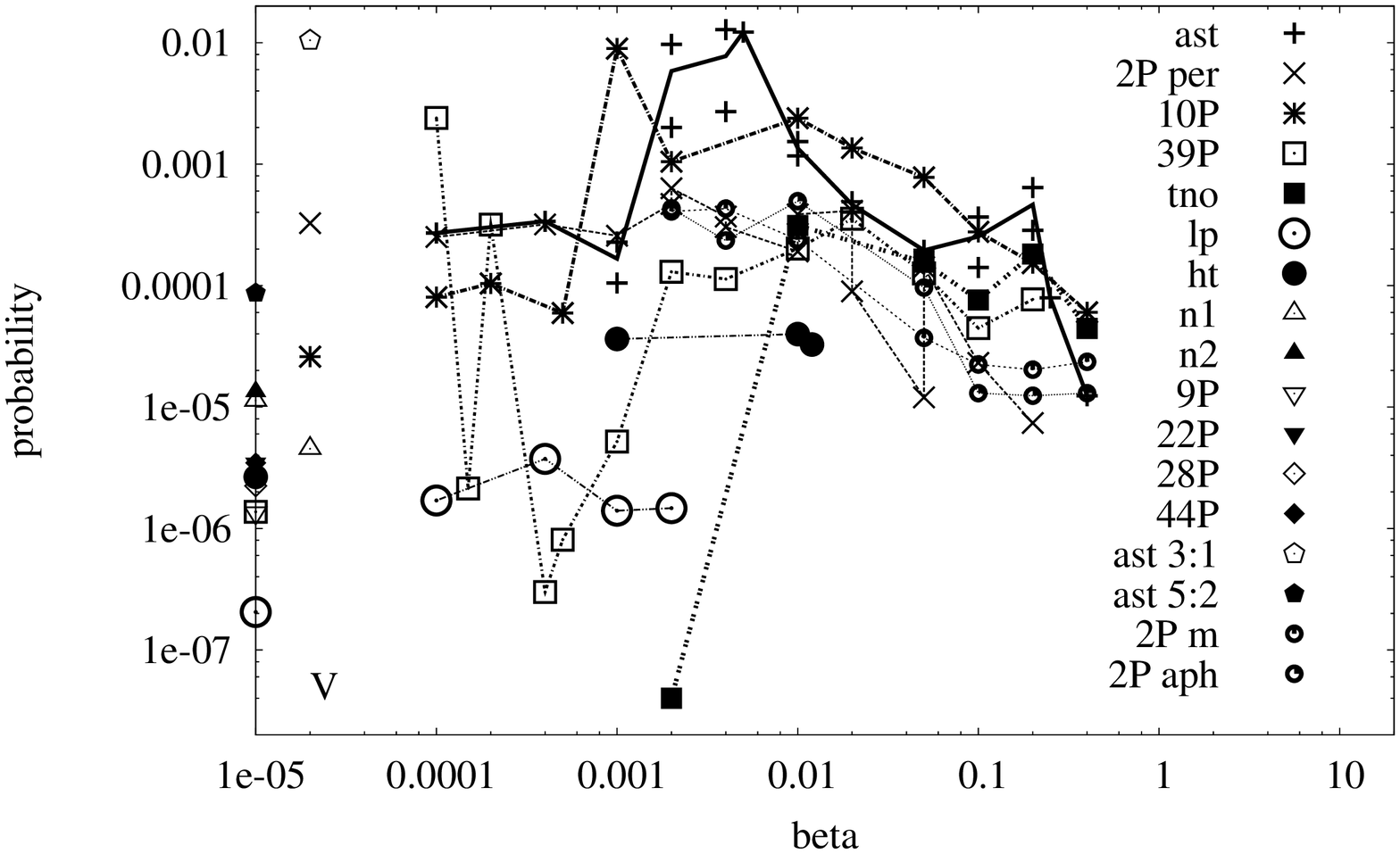}

\label{fig1}
\end{figure}

\begin{figure}[]
\begin{center}
\includegraphics[width=5in]{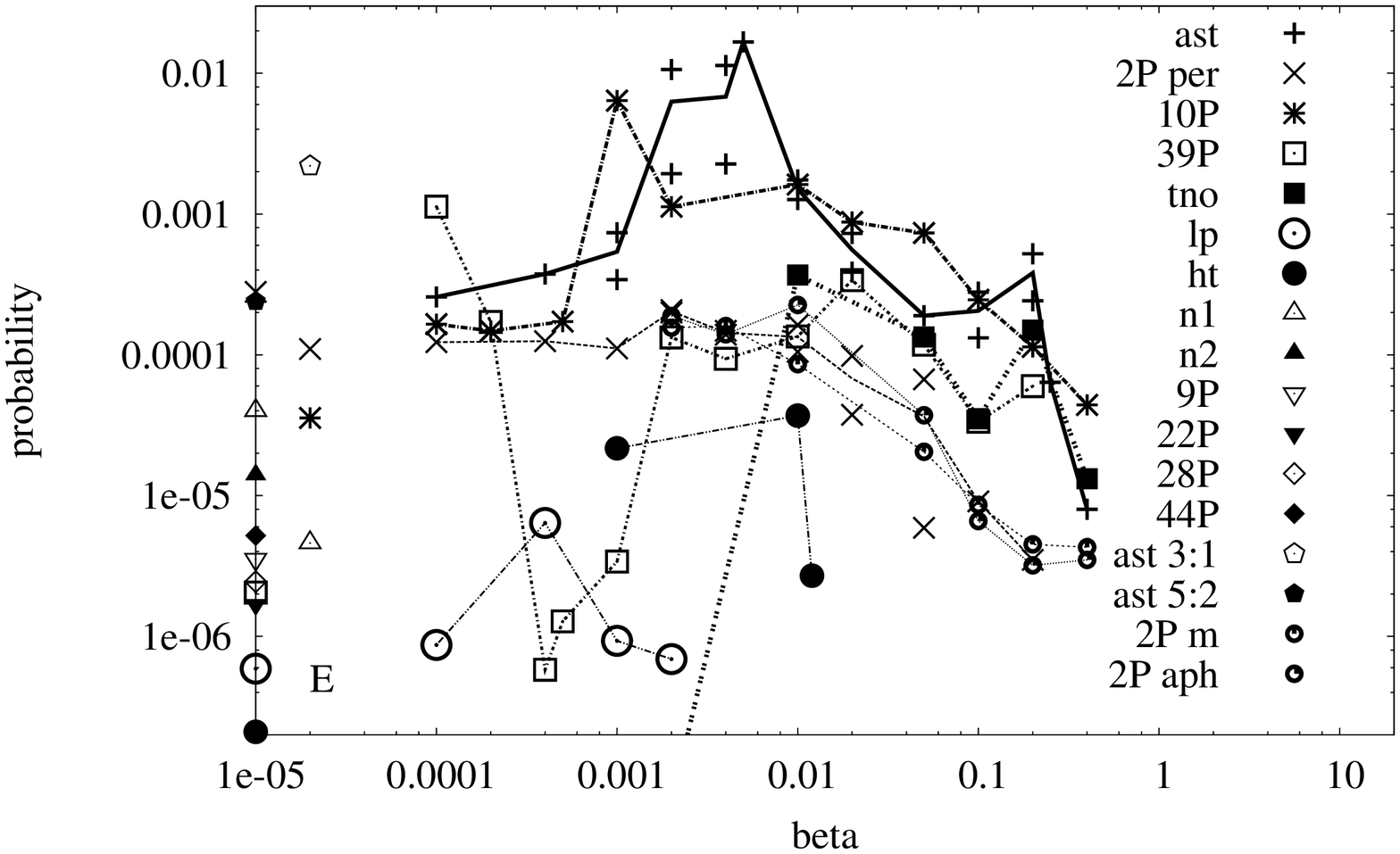}
\includegraphics[width=5in]{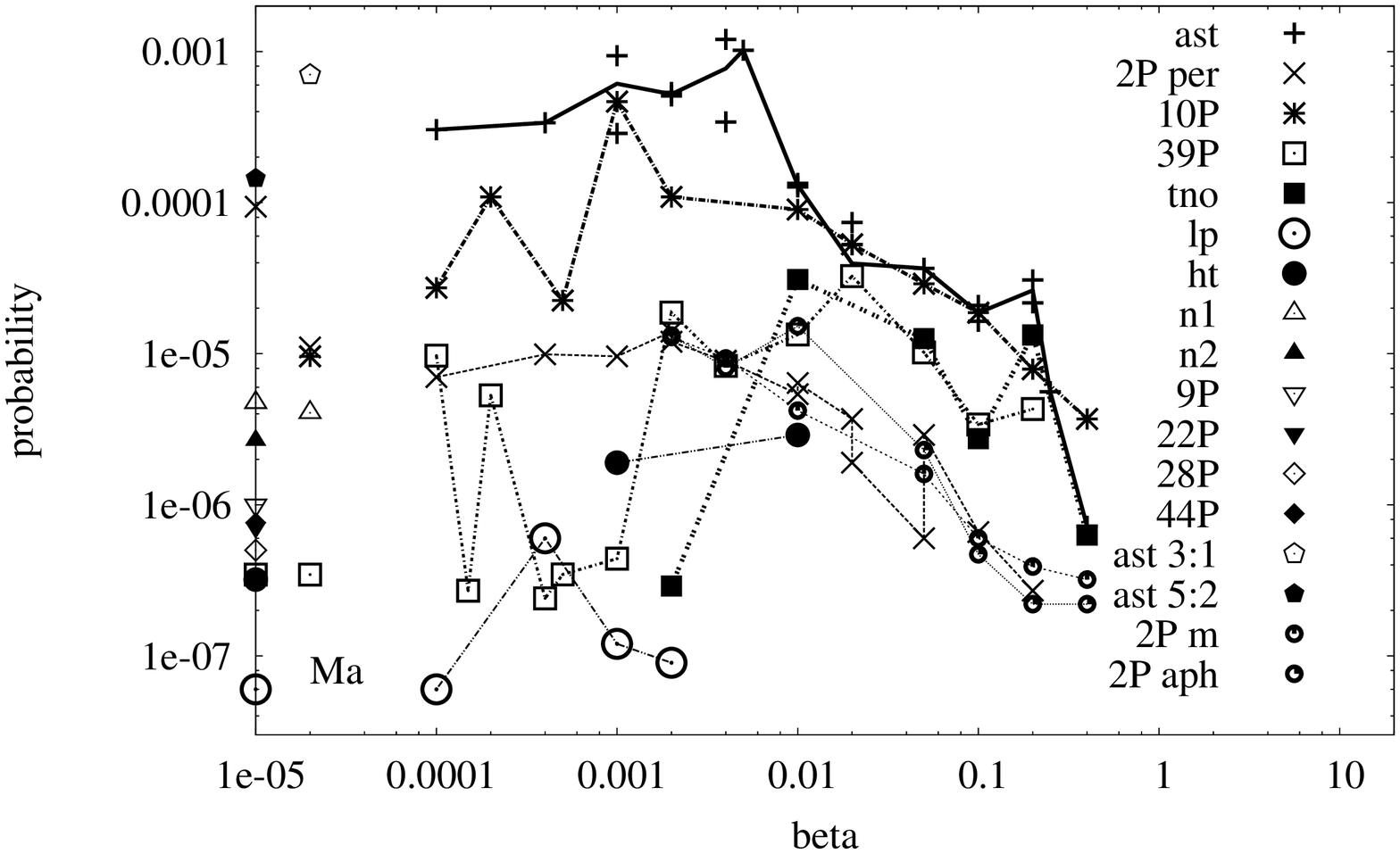}
\includegraphics[width=5in]{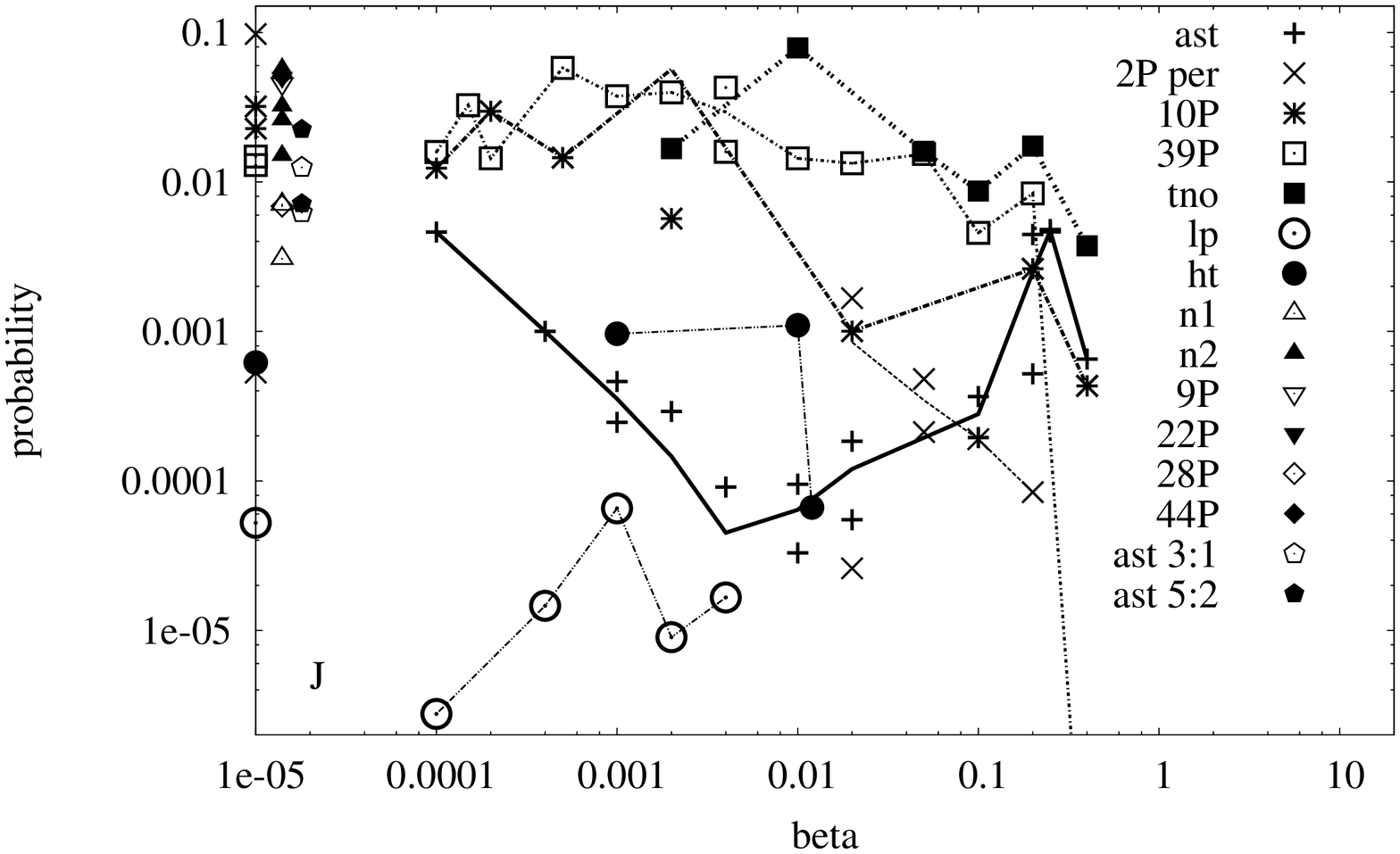}
\label{fig1}
\end{center}
\end{figure}
\begin{figure}[]
\begin{center}
\includegraphics[width=5in]{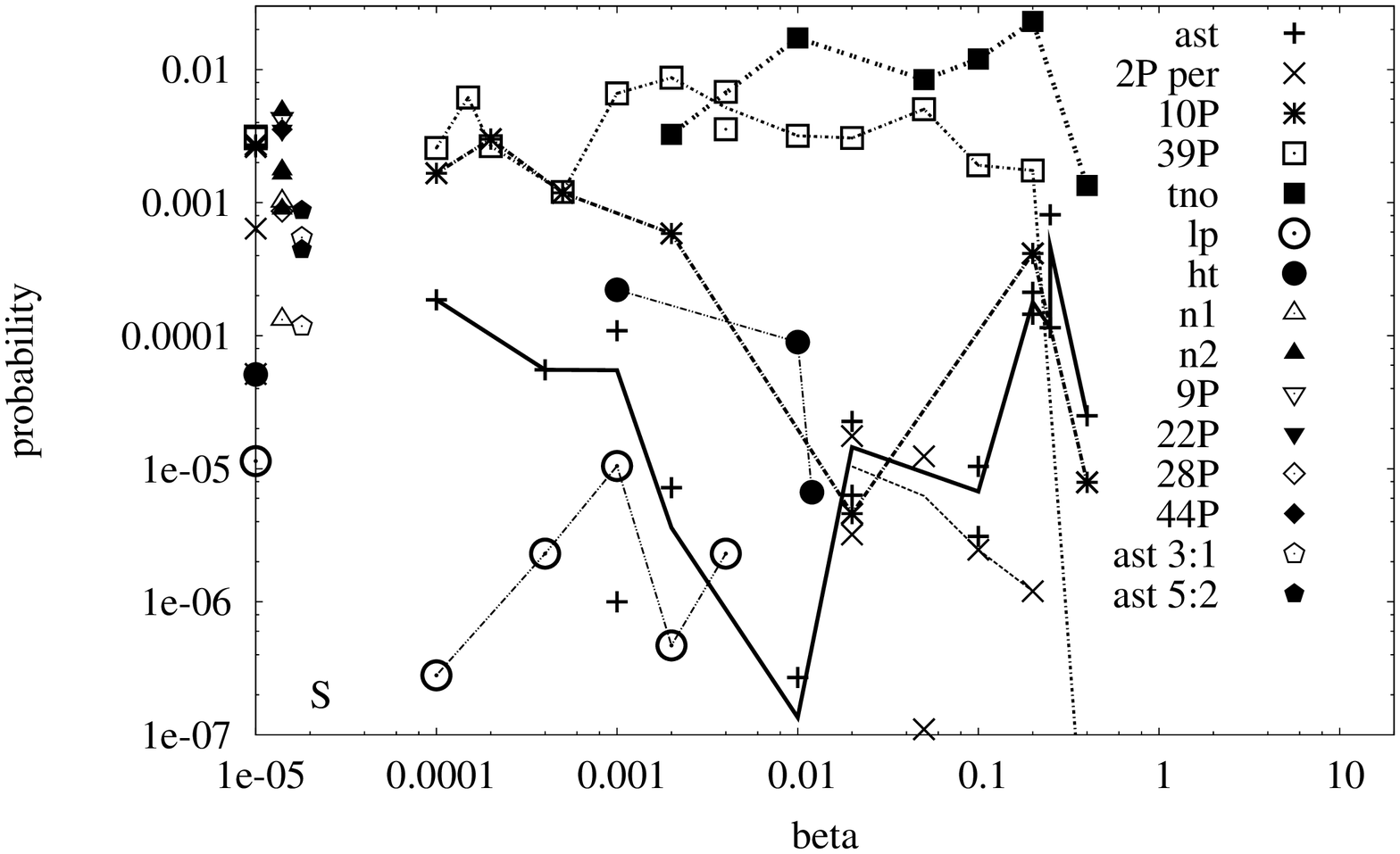}
\includegraphics[width=5in]{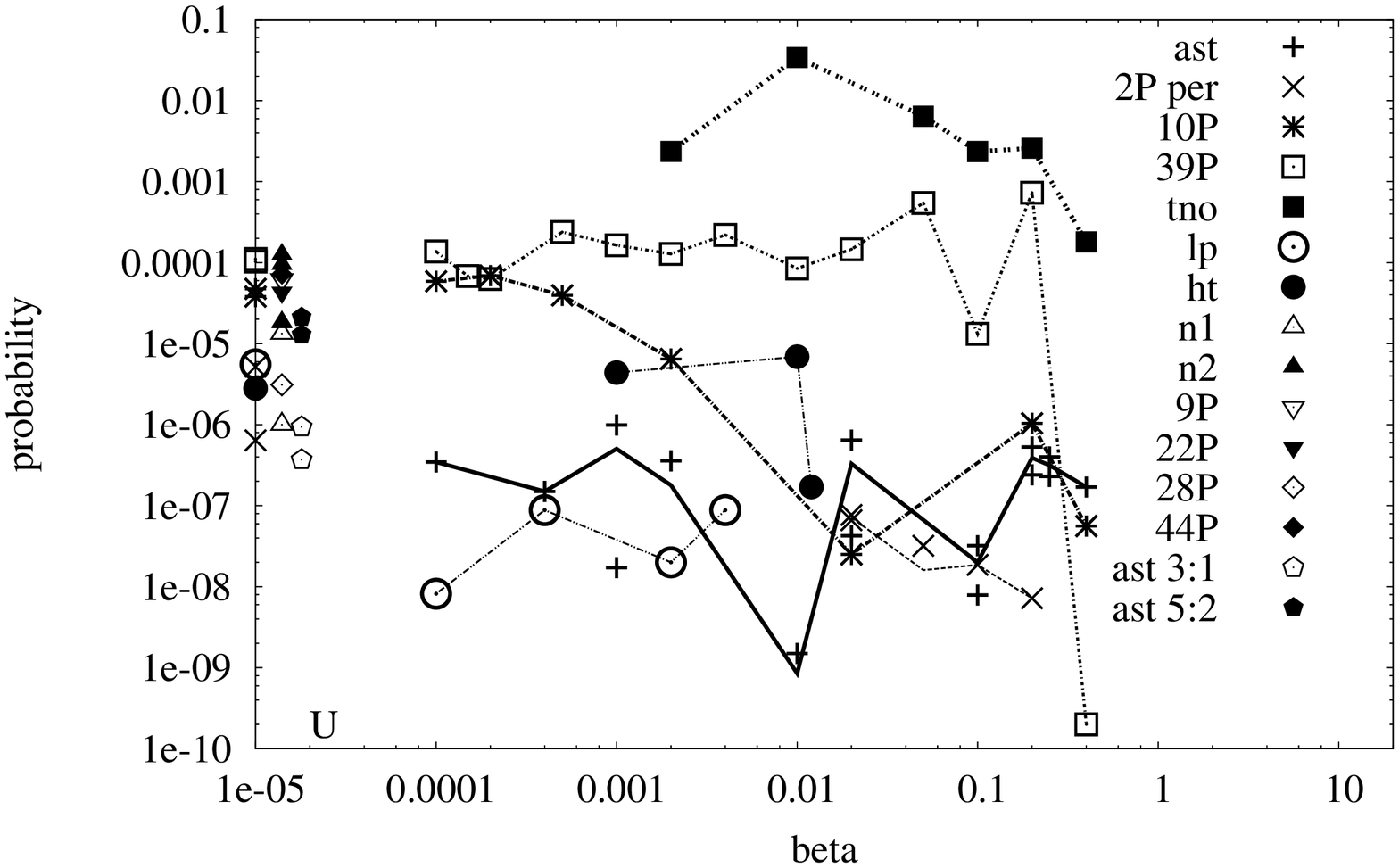}
\includegraphics[width=5in]{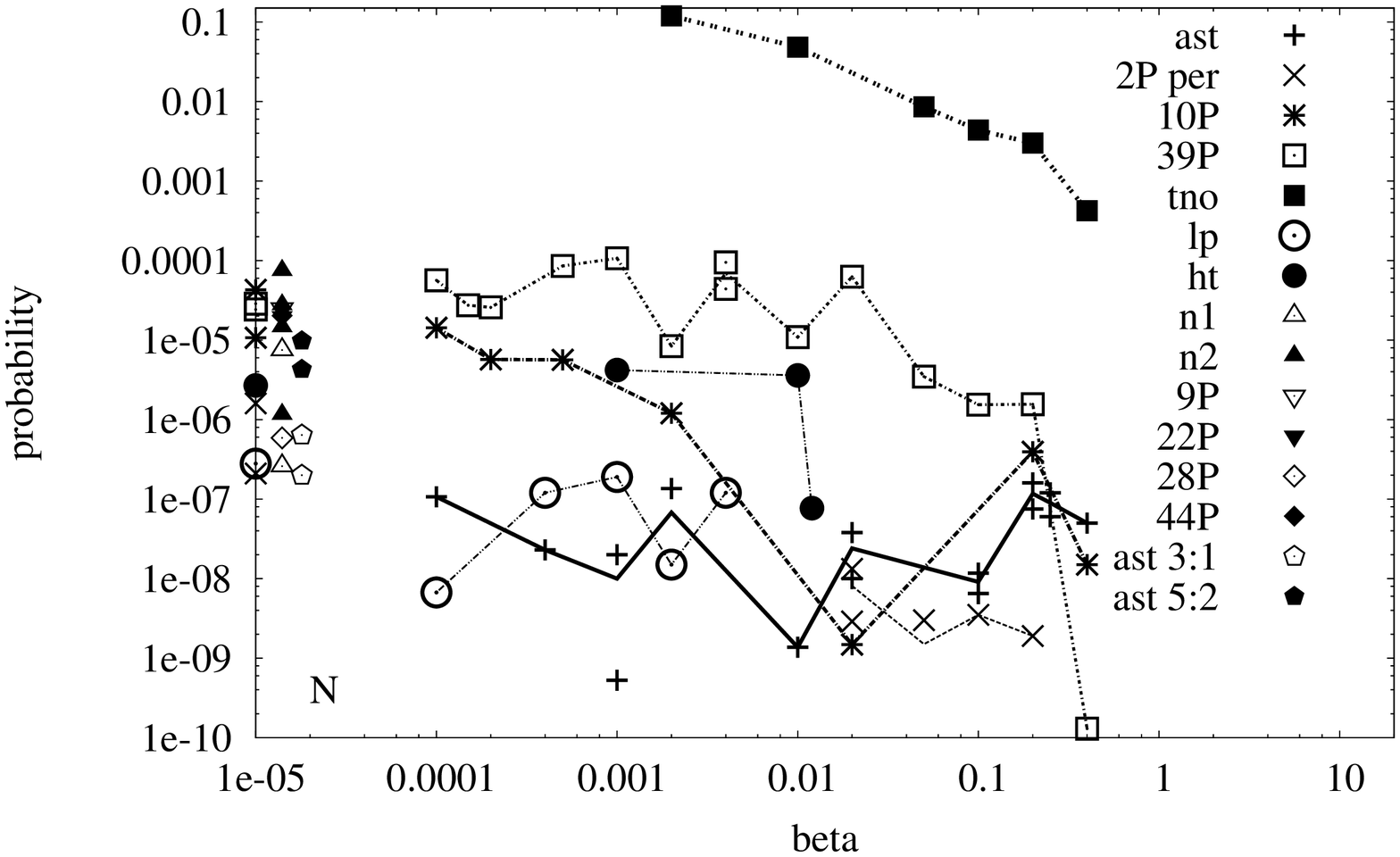}
\label{fig1}
\end{center}
\end{figure}


\begin{thebibliography}{}

\bibitem[11]{11} Drake, M., \& Campins, H.  2006, \textit{Asteroids, Comets, \& Meteors}, 
IAU Symp. 229, 
381

\bibitem[12]{12}
Genda, H., \& Icoma, M.  2008,
\textit{Icarus}, 194, 42

\bibitem[7]{7} Ipatov, S.I. 1987, 
\textit{Earth, Moon, \& Planets}, 39, 101

\bibitem[37]{37}
Ipatov, S.I. 1988, 
\textit{Sov. Astron.}, 65, 1075

\bibitem[8]{8} Ipatov, S.I. 1993, 
\textit{Solar System Res.}, 27, 65

\bibitem[9]{9}  Ipatov, S.I. 2001, 
\textit {Adv. Space Res.}, 28, 1107

\bibitem[1]{1} Ipatov, S.I., \& Mather, J.C. 2003, 
\textit{Earth, Moon, \& Planets}, 92, 89

\bibitem[2]{2}  Ipatov, S.I., \& Mather, J.C. 2004a, 
\textit{Annals of New York Acad. Sci.}, 1017, 46

\bibitem[4]{4}  Ipatov, S.I., \& Mather, J.C. 2004b, 
\textit{Adv. Space Res.}, 33, 1524

\bibitem[5]{5}  Ipatov, S.I., \& Mather, J.C. 2006, 
\textit{Adv. Space Res.}, 37, 126

\bibitem[5]{5}  Ipatov, S.I., \& Mather, J.C. 2007, Proc. IAU Symp. 236 
\textit{Near-Earth Objects, Our Celestial Neighbors: Opportunity and Risk} 
55

\bibitem[3]{3} Ipatov, S.I., Mather, J.C.,  \& Taylor, P.A. 2004, 
\textit{Annals of New York Acad. Sci.}, 1017, 66

\bibitem[26]{26} 
Levison, H.F., \& Duncan, M.J. 1994, 
\textit{Icarus}, 108, 18

\bibitem[27]{27} 
Levison, H.F., Dones, L., Chapman, C.R., Stern, S.A., Duncan, M.J., \& Zahnle, K.  2001, 
\textit{Icarus},  151, 286

\bibitem[13]{13}
Lunine, J.I., Chambers, J., Morbidelli, A., \& Leshin, L.A. 2003, 
\textit{Icarus},  165, 1


\bibitem[25]{25}  Marov, M.Ya., \& Ipatov, S.I. 2006, 
\textit{Solar Syst. Res.}, 39, 374
 
\bibitem[26]{26} Morbidelli, A., Chambers, J., Lunine, J.I., Petit, J.M., Robert, F., Valsecchi, G.B., \& Cyr, K.E.  2000, 
\textit{Meteoritics \& Planet. Sci.},  35, 1309

\bibitem[28]{28} Petit, J.-M., Morbidelli, A., \& Chambers, J. 2001, 
\textit{Icarus}, 153, 338

\end{thebibliography}
\end{document}